\documentclass[notitlepage, 
noruninaddress,
bibnotes,
prb,5pt, 
twocolumn,
showpacs,preprintnumbers,superscriptaddress,
amsmath,amssymb,floatfix]{revtex4-1}
\usepackage{amsmath}
\usepackage{graphicx}
\usepackage{ dsfont }
\usepackage{dcolumn}
\usepackage{bm}
\usepackage{color}
\usepackage{xcolor}
\usepackage{url}
\usepackage{tabularx}
\usepackage{array}
\usepackage{booktabs}
\usepackage{comment}
\usepackage{hyperref}
\usepackage{footnote}
\usepackage{lipsum}
\usepackage[normalem]{ulem}

\newcounter{firstbib}

\renewcommand{\onlinecite}[1]{\nocite{#1}\citenum{#1}} 
\usepackage{titlesec}
\newcommand*{\justifyheading}{\raggedright}
\titleformat{\section}
  {\normalfont\bfseries\justifyheading}{\thesection}{1em}{}  
\usepackage[singlelinecheck=false, justification=RaggedRight, format=plain]{caption}
\DeclareCaptionLabelSeparator{bar}{ $\boldsymbol{\mid}$ }
\newcommand{\thesupplementbibliography}{\thebibliography}
\usepackage{etoolbox}
\makeatletter
\apptocmd{\thesupplementbibliography}{\global\c@NAT@ctr 30\relax}{}{}
\makeatother
\begin{document}

\title{Magnetic skyrmion braids}

\author{Fengshan~Zheng}
\email{f.zheng@fz-juelich.de}
\affiliation{Ernst Ruska-Centre for Microscopy and Spectroscopy with Electrons and Peter Gr\"unberg Institute, Forschungszentrum J\"ulich, 52425 J\"ulich, Germany}

 \author{Filipp~N.~Rybakov}
 \email{prybakov@kth.se}
 \affiliation{Department of Physics, KTH-Royal Institute of Technology, Stockholm, SE-10691 Sweden}

\author{Nikolai~S.~Kiselev}
\email{n.kiselev@fz-juelich.de}
\affiliation{Peter Gr\"unberg Institute and Institute for Advanced Simulation, Forschungszentrum J\"ulich and JARA, 52425 J\"ulich, Germany}

\author{Dongsheng Song}
\affiliation{Ernst Ruska-Centre for Microscopy and Spectroscopy with Electrons and Peter Gr\"unberg Institute, Forschungszentrum J\"ulich, 52425 J\"ulich, Germany}
\affiliation{Information Materials and Intelligent Sensing Laboratory of Anhui Province, Key Laboratory of Structure and Functional Regulation of Hybrid Materials of Ministry of Education, Institutes of Physical Science and Information Technology, Anhui University, Hefei 230601, China}

\author{Andr\'as~Kov\'acs}
\affiliation{Ernst Ruska-Centre for Microscopy and Spectroscopy with Electrons and Peter Gr\"unberg Institute, Forschungszentrum J\"ulich, 52425 J\"ulich, Germany}

\author{Haifeng~Du}
\affiliation{The Anhui Key Laboratory of Condensed Matter Physics at Extreme Conditions, High Magnetic Field Laboratory, Chinese Academy of Science (CAS), Hefei, Anhui Province 230031, China}
 
\author{Stefan~Bl\"ugel}
 \affiliation{Peter Gr\"unberg Institute and Institute for Advanced Simulation, Forschungszentrum J\"ulich and JARA, 52425 J\"ulich, Germany}

\author{Rafal~E.~Dunin-Borkowski}
 \affiliation{Ernst Ruska-Centre for Microscopy and Spectroscopy with Electrons and Peter Gr\"unberg Institute, Forschungszentrum J\"ulich, 52425 J\"ulich, Germany}

\date{\today}

\maketitle


\section*{Abstract}

\textbf{
Filamentary textures can take the form of braided, rope-like superstructures in nonlinear media such as plasmas and superfluids~\cite{Parker1, Cirtain2013, PhysRevLett.96.215302}.
The formation of similar superstructures in solids has been predicted, for example from flux lines in superconductors~\cite{PhysRevLett.60.1973}. However, their observation has proved challenging~\cite{PhysRevLett.92.157002, Reichhardt2009}.
Here, we use electron microscopy and numerical methods to reveal braided superstructures of magnetic skyrmions~\cite{Bogdanov_89, skyrmionics_roadmap_review_2020, Bogdanov_review_2020, Tokura_review_2020} in thin crystals of B20-type FeGe.
Their discovery opens the door to applications of rich topological landscapes of geometric braids~\cite{Artin_1947, braids_textbook1} in magnetic solids.
}
%


\section*{Main}





\begin{figure*}[ht] 
\centering
\includegraphics[width=17cm]{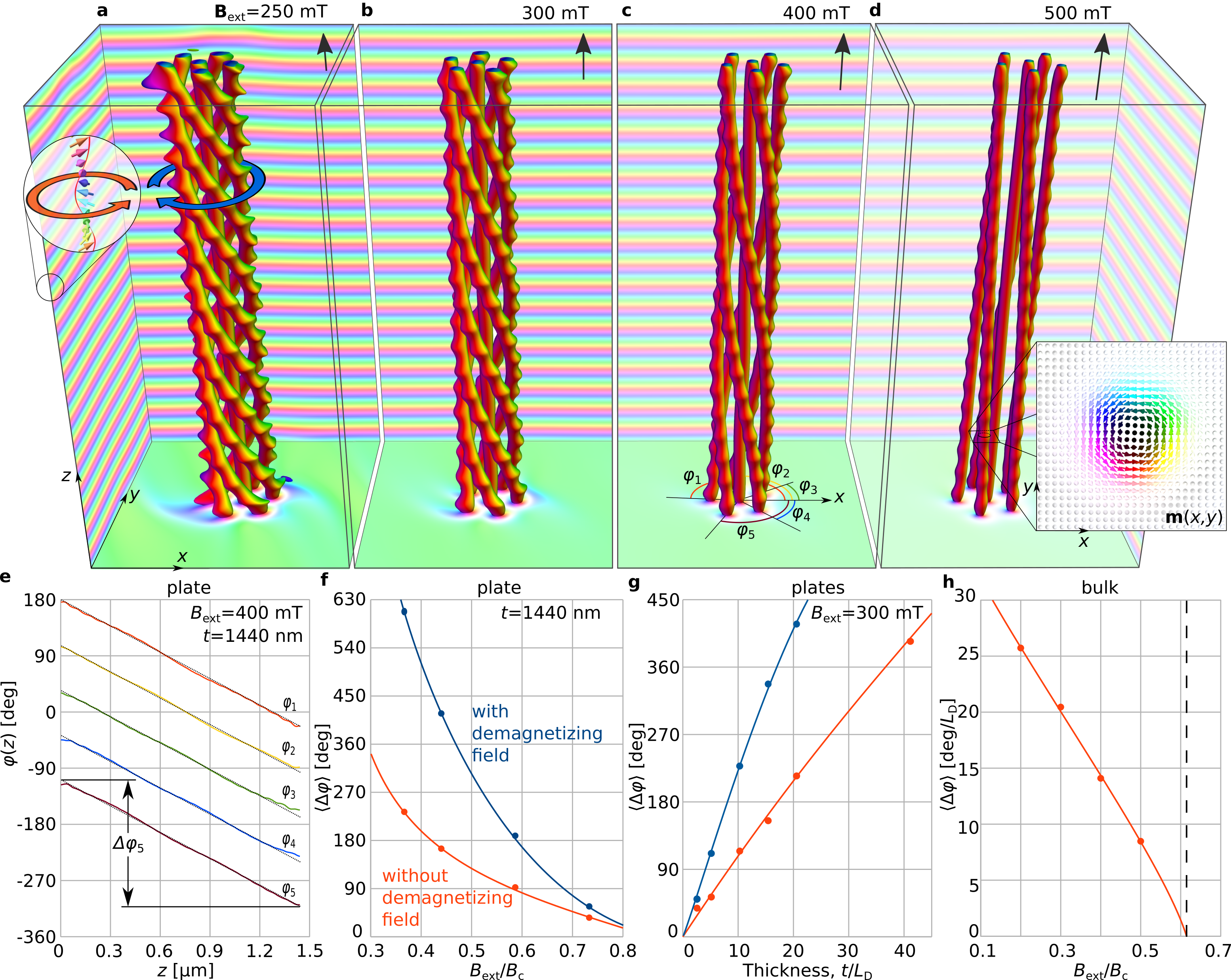}
\caption{\small
\textbf{Skyrmion braid in a chiral magnet.}
\textbf{a}-\textbf{d}, Skyrmion braid comprising six skyrmion strings represented by isosurfaces of ${m_\mathrm{z}=0}$.
The color modulation at the edges of the box indicates the presence of the conical phase, see inset in \textbf{a}.
The equilibrium state for each value of applied magnetic field ($\mathbf{B}_\mathrm{ext}||\hat{\mathbf{e}}_\mathrm{z}$) is found by using energy minimization (see Methods), on the assumption of periodic boundary conditions in the $xy$~plane, free surfaces in the third dimension and a sample thickness of ${t = 1440}$~nm.
\textbf{e}, Dependence of twist angle on distance to the lower surface for the five skyrmions that wind around a central sixth skyrmion.
The angles $\varphi_i(z)$ are measured from the $x$~axis, as indicated in \textbf{c}.
The dotted lines are linear fits for each $\varphi_i(z)$ dependence.
\textbf{f}, Dependence of average twist angle $\langle\Delta\varphi\rangle=\frac{1}{5}\sum_i \left[\varphi_i(0)-\varphi_i(t)\right]$ on applied magnetic field. 
\textbf{g}, Dependence of average twist angle $\langle\Delta\varphi\rangle$ on sample thickness $t$.
\textbf{h}, Dependence of energetically-optimal twist per unit length on applied magnetic field for a bulk sample with periodic boundary conditions in all three dimensions.
The solid circles in \textbf{f}-\textbf{h} are obtained from numerical calculations, while the lines are used to guide the eye.
}
\label{Fig_1}
\end{figure*}

Magnetic skyrmions have been observed over a wide range of temperatures and applied magnetic fields in Ge-based and Si-based alloys with B20-type crystal structures, such as FeGe~\cite{Yu_11, Kovacs_17, Du_18, Yu_18}, MnSi~\cite{Yu_15}, Fe$_{1-x}$Co$_x$Si~\cite{Yu_10, Park_14} and others~\cite{Shibata_13}.
These skyrmions are topologically nontrivial textures of the magnetization unit vector field ${\mathbf{m}(\mathbf{r})}$, whose filamentary structures resemble vortex-like strings or tubes, with typical diameters of tens of nanometers.
The length of such a string is assumed to be limited only by the shape and size of the sample~\cite{Milde}, with direct observations confirming the formation of micrometer-long skyrmion strings~\cite{Birch_2020, Yu_Masell_Tokura_2020}.
Previous studies have suggested that magnetic skyrmion strings are (nearly) straight~\cite{Rybakov_13, Milde, Rybakov_15, Leonov_2016, Hayley_2019, Birch_2020, Yu_Masell_Tokura_2020}.
In contrast, here we show that skyrmion strings are able to twist around one another to form superstructures that we refer to as \textit{skyrmion braids} (Fig.~\ref{Fig_1}). 

\begin{figure*}[ht] 
\centering
\includegraphics[width=18cm]{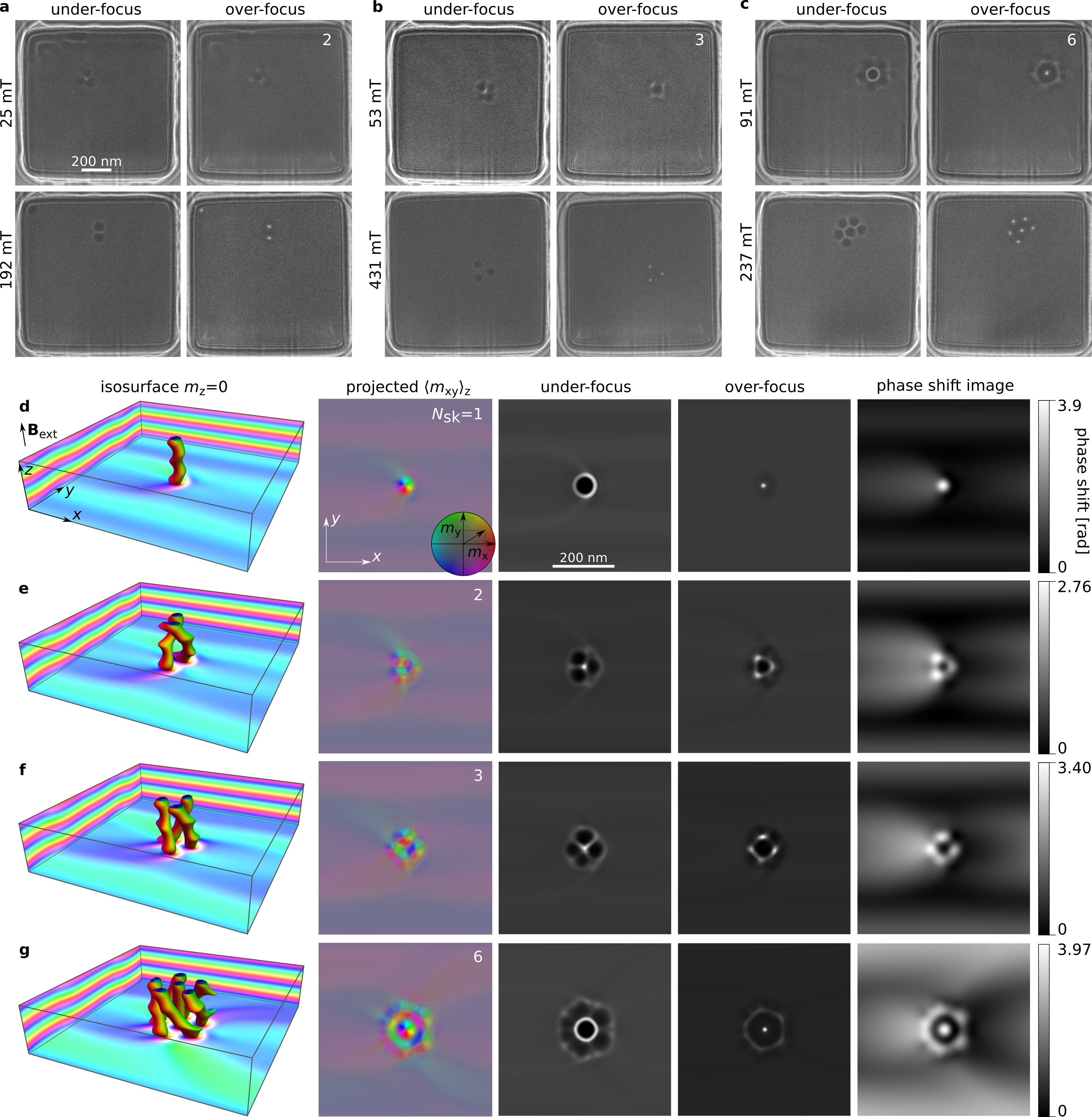}
\caption{\small
\textbf{Skyrmion braids in an extended FeGe plate of thickness 180~nm.}
\textbf{a}-\textbf{c}, 
Experimental Lorentz TEM images of twisted skyrmions -- skyrmion braids (upper row) and untwisted (lower row) skyrmions recorded at two different magnetic fields in sample~S1 at 95~K.
%
%
\textbf{d}-\textbf{g}, Theoretical results for ${B_\text{ext}=275}$~mT, assuming periodic boundary conditions in the $xy$~plane and free surfaces in the third dimension.
From left to right, each row shows: an equilibrium state represented by isosurfaces of $m_\mathrm{z}=0$; in-plane magnetization $\langle \mathbf{m}_\mathrm{xy}\rangle_\mathrm{z}$ averaged over the thickness of the plate; under-focus and over-focus Lorentz TEM images; electron optical phase image.
The index in the top right corner in \textbf{a}-\textbf{c} and in the simulated images of $\langle \mathbf{m}_\mathrm{xy}\rangle_\mathrm{z}$ indicates the number of skyrmion strings $N_\mathrm{sk}$ in the braid.
An isolated skyrmion string is shown in \textbf{d} for comparison with skyrmion braids in \textbf{e}-\textbf{g}. 
}
\label{Fig_2}
\end{figure*}

\begin{figure*}[ht] 
\centering
\includegraphics[width=18cm]{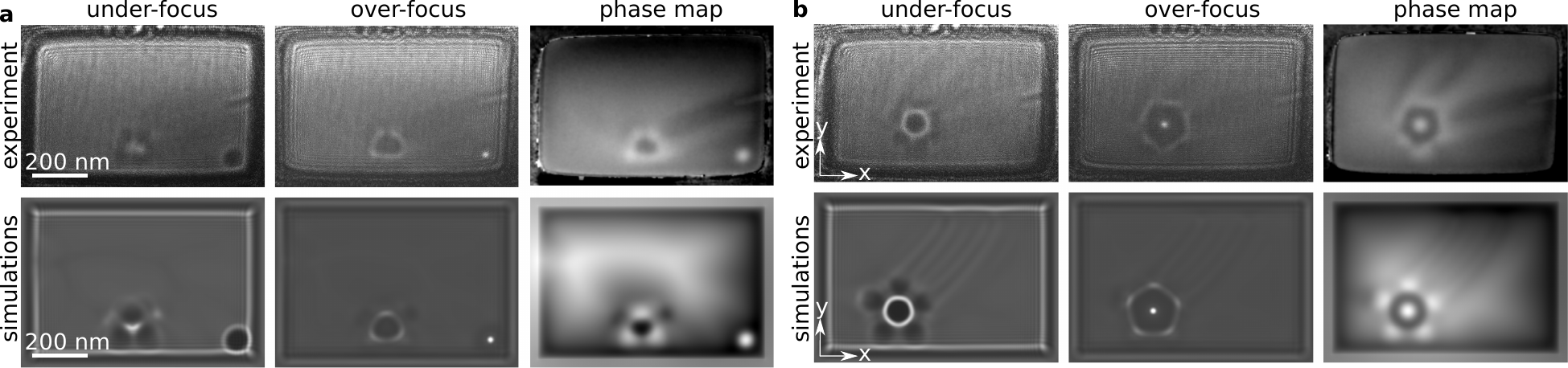}
\caption{\small
\textbf{Comparison between experimental and simulated images of skyrmion braids.}
\textbf{a}-\textbf{b},~Magnetic images of skyrmion braids containing 3 and 6 skyrmion strings, respectively.
%
%
The upper panel shows experimental under-focus and over-focus Lorentz TEM images and electron optical phase shift images recorded using off-axis electron holography from sample~S2 at 95~K.
The lower panel shows corresponding simulated Lorentz TEM images and phase shift images.
The perpendicular applied magnetic field in \textbf{a} and \textbf{b} is 218 and 187~mT, respectively.
The high contrast feature in the lower right corner in \textbf{a} is an isolated skyrmion string attached to the edge.
The sample size, external magnetic field, defocus and accelerating voltage are the same for the experimental and  simulated images.
%
%
}
\label{Fig_3}
\end{figure*}

We begin with a theoretical description, which is based on a micromagnetic approach and a standard model of cubic helimagnets (chiral magnets), such as FeGe (see Methods).
Static equilibrium in such a system results from a delicate balance between Heisenberg exchange, Zeeman interaction, chiral Dzyaloshinskii-Moriya interaction~\cite{Dzyaloshinskii,Moriya} (DMI) and dipole-dipole interaction.
The equilibrium state is typically a conical spin spiral, whose modulation axis is parallel to the applied field $\mathbf{B}_\text{ext}$ (inset to Fig.~\ref{Fig_1}\textbf{a}).
The period of such a spiral $L_\mathrm{D}$ is nearly constant for each material, taking a value of $\sim70$~nm for FeGe~\cite{Lebech}.
The conical phase can be regarded as a natural ``vacuum''~(background), in which clusters and isolated strings of skyrmions are embedded~\cite{Rybakov_15, Leonov_2016}. 
%
%

We use energy minimization (see Methods) to show that interacting skyrmion strings can lower their total energy by twisting into braids.
A representative example of such a skyrmion braid is shown in Fig.~\ref{Fig_1} for an energetically favorable configuration of five skyrmion strings, which wind around a sixth straight string at their center.
Such a winding braid forms spontaneously from an initial configuration of six straight skyrmion strings (see Methods and Extended Data Video~1).
A skyrmion braid has one energetically-preferable chirality, which depends on the sign of the DMI constant $D$.
Without loss of generality, we assume below that $D>0$.
The chirality of the twist is opposite to that of the magnetization of the surrounding cone phase, as indicated using blue and red arrows, respectively, in Fig.~\ref{Fig_1}\textbf{a}.
Although this behavior may appear to be counter-intuitive, along any line parallel to the $z$~axis the spins have the same sense of rotation about $\hat{\mathbf{e}}_\mathrm{z}$ as for the conical spiral.
The twisting or untwisting of a skyrmion braid by an external magnetic field, as illustrated in Figs~\ref{Fig_1}\textbf{a}-\textbf{d}, is fully reversible (see Extended Data Video~1).

In order to quantify the degree of twist, we trace the position of each skyrmion by its center, where the magnetization is antiparallel to $\mathbf{B}_\text{ext}$.
We define the twist angle $\varphi_i(z)$ for the $i$\textsuperscript{th} skyrmion in each $z$~plane as the angle between the $x$~axis and the line connecting the central and $i$\textsuperscript{th} skyrmion (Fig.~\ref{Fig_1}\textbf{c}).
The $\varphi_i(z)$ curves shown in Fig.~\ref{Fig_1}\textbf{e} are almost linear and have identical slopes, with small deviations close to the free surfaces due to the chiral surface twist~\cite{Rybakov_13}.
A linear fit to each plot of $\varphi_i(z)$ can be used to estimate the total twist angle $\Delta\varphi_i$, by which the strings rotate from the lower to the upper surface.

Figure~\ref{Fig_1}\textbf{f} shows the dependence of the average twist angle $\langle\Delta\varphi\rangle=\frac{1}{5}\sum_i \Delta\varphi_i$ on the applied magnetic field, both with and without consideration of demagnetizing fields, to illustrate the role of dipole-dipole interactions.
The twist angle is reduced by a factor of $\sim$2 when demagnetizing fields are neglected. This difference decreases with increasing $B_\mathrm{ext}$.
The applied magnetic field is shown in reduced units relative to the critical field $B_\mathrm{c}$ in Fig.~\ref{Fig_1}\textbf{f}, in order to facilitate a comparison between the results obtained with and without demagnetizing fields.
In both cases, $B_\mathrm{c}$ is the phase transition field between conical and saturated ferromagnetic states (see Methods).
Figure~\ref{Fig_1}\textbf{g} shows the dependence of the twist angle $\langle\Delta\varphi\rangle$ on sample thickness $t$ at fixed $B_\mathrm{ext}$.
Although dipole-dipole interactions increase the degree of twist significantly, they are not necessary for the stability of skyrmion braids.

The quasilinear dependence of $\varphi_i(z)$ on $z$ in Fig.~\ref{Fig_1}\textbf{e} and the monotonic dependence of $\langle\Delta\varphi\rangle(t)$ on $t$ in Fig.~\ref{Fig_1}\textbf{g} suggest that the volume and the free surfaces both play important roles in the formation of skyrmion braids.
Calculations performed for a bulk sample with periodic boundary conditions show that a six-string skyrmion braid is energetically more favourable than a cluster of straight strings in fields below ${B_\text{ext}\approx 0.61 B_\text{c}}$ (see the vertical dashed line in Fig.~\ref{Fig_1}\textbf{h}).
In a thin plate, the same braid is stable above this value of applied field (Fig.~\ref{Fig_1}\textbf{f}), suggesting that the surface effects may additionally enhance the braids stability.
%

The dependence of average twist angle $\langle\Delta\varphi\rangle$ on plate thickness shown in Fig.~\ref{Fig_1}\textbf{g} suggests that the helical twist of skyrmion strings should be observable in a thin sample, for which ${t\sim L_\mathrm{D}}$.
This criterion is important for observations using transmission electron microscopy (TEM), for which an electron-transparent sample is required.
For Fe$_{0.5}$Co$_{0.5}$Si, with ${L_\mathrm{D}\sim 90}$~nm, the thickness below which a sample is electron-transparent at an accelerating voltage of 300~kV is ${\sim 3.3 L_\mathrm{D}}$, corresponding to $\sim$300~nm~\cite{Park_14}.
This value is assumed to be approximately the same for FeGe. 

We studied three high-quality electron-transparent FeGe lamellae with thicknesses of $\sim180$~nm (${\sim 2.6L_\mathrm{D}}$), which is large enough for the formation of skyrmion braids but small enough for magnetic imaging in the TEM.
The lateral dimensions of the samples were: 1~$\mu$m~$\times$~1~$\mu$m~(S1); 800~nm~$\times$~540~nm~(S2); 6~$\mu$m~$\times$~7~$\mu$m~(S3).
We observed skyrmion braids in all three samples.
Below, we provide representative results from samples S1 and S2. Other results are included in Extended Data Figs~2-9 and Videos~3 and 4.

For experimental observations of skyrmion braids, we designed the following protocol.
First, we performed magnetization reversal cycles, until a desired number of skyrmion strings had nucleated in the sample. Multiple cycles were required due to the probabilistic character of skyrmion nucleation.
In order to avoid the interaction of skyrmions with the edges of the sample, the strength of the external magnetic field was increased to move the skyrmions towards the center of the sample.
We then gradually decreased the applied magnetic field until the contrast of the straight skyrmion strings changed to those expected for braids.
Although this is not a unique protocol for nucleating skyrmion braids (see, \emph{e.g.}, Extended Data Fig.~3), it results in high reproducibility of the results (see Extended Data Video.~4).
We did not observe skyrmion braids during in-field cooling of the samples.

Figure~\ref{Fig_2} shows experimental Lorentz TEM images (\textbf{a}-\textbf{c}) and theoretical results (\textbf{d}-\textbf{g}) for short skyrmion braids.
Here we compare only the Lorentz TEM images. Representative experimental phase shift images for sample S1 are provided in Extended Data Fig. 2.
One can see good agreement between the experimental and simulated images.
The key feature of these magnetic images of skyrmion braids is the appearance of the ring-like pattern and the significant reduction of the contrast in comparison to straight skyrmions, see the lower row in Figs~\ref{Fig_2}\textbf{a}-\textbf{c}.
The correlation between the magnetic images and average in-plane magnetization components $\langle\mathbf{m}_\mathrm{xy}\rangle_\mathrm{z}=t^{-1}\int\mathbf{m}_\mathrm{xy}dz$ in Figs~\ref{Fig_2}\textbf{d}-\textbf{g} arises because, to a first approximation, one may assume that the magnetic field follows the $\mathbf{m}$ vector.
It should be noted that the electron beam is perpendicular to the plate and only sensitive to $xy$~component of the magnetic field within and around the sample.
In sample~S1 (and similarly in sample~S3 -- see Extended Data Fig.~9), the typical applied fields that were found to stabilize braids were different from theoretical values.
This discrepancy can be attributed to minor damages from the fabrication process~\cite{Kato_04}, including the formation of thin amorphous magnetic surface layers~\cite{Suran_76(1), Suran_76(2), Mangin_78}.
In contrast, excellent agreement with theory was achieved for sample~S2, which was also used in Ref.~\onlinecite{Zheng_18}.

Figure~\ref{Fig_3} shows experimental and theoretical images for sample S2 (see also Extended Data Fig.~4).
Figure~\ref{Fig_3}\textbf{a} shows a three-string skyrmion braid.
Figure~\ref{Fig_3}\textbf{b} shows a skyrmion braid comprising six strings in a pentagonal pattern, similar to those shown in Figs~\ref{Fig_2}\textbf{c} and \ref{Fig_2}\textbf{g}.
As a result of its smaller size, electron phase shift images could be recorded from this sample using off-axis electron holography (see Methods).
Excellent agreement is obtained between experiment and theory for both under-focus and over-focus Lorentz TEM images and electron phase shift images, providing confidence in the interpretation of the experimental data.
The skyrmion braids can be stabilized over a wide temperature range, as shown in the form of representative magnetic images in Extended Data Fig.~8 for sample~S1 at 120~K and Extended Data Fig.~6 for sample~S2 at 170~K.

%
Because of the magnetization inhomogeneities along the skyrmion braids, their electromagnetic properties are expected to differ from those of straight skyrmion strings.
Electrons passing along the braids accumulate a Berry phase and experience emergent electric and magnetic fields modulated with the thickness of the sample and quantized by the number of skyrmion strings in the braid. This in turn gives rise to magnetoresistive and transport phenomena which would not be observed for straight skyrmion strings.
A promising system for demonstrating these effects is a magnetic nanowire~\cite{Liang_Song_2015, Stolt_Song_2017}, in which a skyrmion braid can occupy the entire volume.
In larger samples, skyrmion braids are expected to differ in structure and close-packing form~\cite{Olsen_2010, Bohr_2011}.

A skyrmion braid represents a geometric braid~\cite{braids_textbook2} when the ends of the skyrmion strings are fixed on the sample surface due to pinning or surface engineering and is then topologically protected and cannot be unwound.
%
The latter is guaranteed by robustness of skyrmion strings resembling strides~\cite{braids_textbook1} of a geometric braid.
Such skyrmion braids can be described in terms of an Artin braid group $\mathcal{B}_N$. 
In this sense, skyrmion braids are a three-dimensional analogues of anyons~\cite{anyons} in (2+1)-dimensional space-time.
%


In summary, we have discovered novel magnetic superstructures, which we term skyrmion braids, using electron microscopy and micromagnetic calculations.
The generality of our theoretical approach suggests that superstructures of skyrmion strings that wind around one another can be formed in all noncentrosymmetric cubic magnets, offering new perspectives for studies and applications of the anomalous (topological) Hall effect~\cite{Ye_1999, Nagaosa_2010, Liang_Song_2015}, magnetic resonance, spin waves on skyrmion strings~\cite{Seki_spin_waves_2020, Xing_Braun_spin_waves_2020} and magnetization dynamics driven by currents~\cite{Liang_Song_2015}.
%
%
%




\renewcommand{\bibname}{Literary works}

\textbf{Acknowledgments.}
The authors are grateful to S.~Wang, R.~Jiang and T.~Denneulin for asistance with TEM sample preparation, and for funding to the European Research Council under the European Union's Horizon 2020 Research and Innovation Programme (Grant No.~856538 - project ``3D MAGiC''; Grant No.~823717 - project ``ESTEEM3''; Grant No.~766970 - project ``Q-SORT''), to the Deutsche Forschungsgemeinschaft (Project-ID 405553726 – TRR~270; Priority Programme SPP 2137; Project No.~403502830), and to the DARPA TEE program through grant MIPR\# HR0011831554. 
F.N.R. was supported by Swedish Research Council Grants 642-2013-7837, 2016-06122 and 2018-03659, and by the G\"{o}ran Gustafsson Foundation for Research in Natural Sciences.
N.S.K. acknowledges financial support from the Deutsche Forschungsgemeinschaft through SPP 2137 "Skyrmionics" Grant No. KI 2078/1-1.


\textbf{Author contributions.}
F.Z., F.N.R. and N.S.K. conceived the project, and contributed equally to the work.
F.Z. and N.S.K. performed TEM experiments and data analysis.
F.N.R. performed simulations, with assistance from N.S.K..
All of the authors discussed the results and contributed to the final manuscript.

\textbf{Competing interests.}
The authors declare no competing interests.

\end{document}